\documentclass[11pt,a4paper,twoside]{article}

\usepackage{amsfonts,amssymb,amsthm}
\usepackage{hyperref}

\usepackage{amsmath}
\usepackage{amsfonts}
\usepackage{amstext}
\usepackage{amsopn}
\usepackage{amsxtra}

\include{diagxy}

\newcommand{\R}{\mathbb{R}}
\newcommand{\C}{\mathbb{C}}
\newcommand{\I}{\mathrm{i}}
\newcommand{\E}{\mathrm{e}}

\newcommand{\N}{\mathbb{N}}
\newcommand{\Z}{\mathbb{Z}}
\newcommand{\T}{\mathbb{T}}
\newcommand{\Hi}{\mathcal{H}}
\newcommand{\B}{\mathcal{B}}
\newcommand{\U}{\mathcal{U}}
\newcommand{\UB}{\mathcal{U}_{\rm B}}
\newcommand{\UZ}{\mathcal{U}_{\rm Z}}

\newcommand{\W}{\mathcal{W}}
\newcommand{\J}{\mathcal{J}}

\newcommand{\Sch}{\mathcal{S}}
\newcommand{\A}{\mathcal{A}}

\newcommand{\Lap}{\Delta}

\newcommand{\ph}{\varphi}

\newcommand{\indexd}{\in\{ 1,\ldots,d \}}

\newcommand{\Ht}{\mathcal{H}_{\tau}}
\newcommand{\Hf}{\mathcal{H}_{\mathrm{f}}}

\newcommand{\Base}{B}
\renewcommand{\t}{\mathcal{T}}

\newcommand{\be}{\begin{equation}}
\newcommand{\ee}{\end{equation}}

\renewcommand{\labelenumi}{{\rm(\roman{enumi})}}

\newtheorem{theorem}{Theorem}

\newtheorem{proposition}[theorem]{Proposition}

\newtheorem*{assumption}{Assumption}

\theoremstyle{definition}

\newcommand{\bb}{\begin{proof}[Beweis]}
\newcommand{\eb}{\end{proof}}

\newcommand{\ta}{\tau}
\newcommand{\la}{\lambda}

\newcommand{\ie}{{\sl i.e.\/ }}

\newcommand{\eg}{{\sl e.g.\/ }}

\def\d{{\partial}}

\def\({\left(}
\def\){\right)}
\def\<{\left\langle}
\def\>{\right\rangle}

\DeclareMathOperator{\Tr}{Tr}
\DeclareMathOperator{\tr}{tr}
\DeclareMathOperator{\Ran}{Ran}

\begin{document}

\title{\LARGE\bf Triviality of  Bloch and\\ Bloch-Dirac bundles}

\author{ \large Gianluca Panati \medskip  \\
\normalsize Zentrum Mathematik and Physik Department,\\
\normalsize Technische Universit\"{a}t M\"{u}nchen, 80290
M\"{u}nchen, Germany   \medskip}
\date{January 16, 2006}
\maketitle

\begin{abstract}
In the framework of the theory of an electron in a periodic potential, we reconsider
the longstanding problem of the existence of smooth and periodic quasi-Bloch
functions, which is shown to be equivalent to the triviality of the Bloch bundle. By
exploiting the time-reversal symmetry of the Hamiltonian and some bundle-theoretic
methods, we show that the problem has a positive answer in any dimension $d \leq 3$,
thus generalizing a previous result by G.~Nenciu. We provide a general formulation
of the result, aiming at the application to the Dirac equation with a periodic
potential and to piezoelectricity.
\end{abstract}




\section{Introduction}

Many relevant properties of crystalline  solids can be understood by the analysis of
Schr\"{o}dinger operators in the form
\begin{equation}\label{Hamiltonian}
    H = - \Delta + V_{\Gamma},
\end{equation}
where the potential $V_{\Gamma}: \R^d \rightarrow \R$ is  periodic with respect to a
lattice $\Gamma \subset \R^d$. Here by \emph{lattice} we mean a maximal subgroup of
the group $(\R^d,+)$, thus $\Gamma \cong \Z^d $. \noindent As realized at the dawn
of quantum mechanics, the analysis of operators in the form \eqref{Hamiltonian} is
greatly simplified by the use of the Bloch-Floquet transform, here denoted as $\UB$.
The advantage of this construction is that the transformed Hamiltonian $\UB \, H \,
\UB^{-1}$ is a fibered operator with respect to a parameter $k \in \T^d$ (called
\emph{crystal momentum} or \emph{Bloch momentum}) and that, under very general
assumptions on $V_{\Gamma}$, each fiber operator $H(k)$ has compact resolvent and
thus pure point spectrum accumulating at infinity.  We label the eigenvalues in
increasing order, \ie $E_0(k) \leq E_1(k) \leq \ldots$ The function $E_n$ is called
the $n$-th \emph{Bloch band}.

\goodbreak

In many applications one is interested in a family of orthogonal projectors $\{P(k)
\}_{k \in \T^d}$, where $P(k)$ is the spectral projector of $H(k)$ corresponding to
a Bloch band, or to a family of Bloch bands, which is separated (pointwise in $k$)
by a gap from the rest of the spectrum. As a particular but important case, one may
consider the spectral projector up to the Fermi energy $E_{\mathrm{F}}$, assuming
that the latter lies in an energy gap for all $k$, a situation which is relevant
when considering the polarization properties of insulators and semiconductors. Since
the map $k \mapsto H(k)$ is periodic and smooth (in the norm-resolvent sense), the
same is true for the map $k \mapsto P(k)$. Moreover, in many cases, $P(k)$ is indeed
analytic over a complex strip $\t_a = \{k \in \C^d: |\, \mathrm{Im}\, k_i \,| <a \}$.
Thus one may raise the following question:

\bigskip

\noindent \textbf{Question (A):} is it possible to choose a system $\{ \ph_{a}(k)
\}_{a=1,\dots,m}$ of eigenfunctions of $P(k)$, spanning $\Ran P(k)$, such that the
maps $k \mapsto \ph_a(k)$ are {smooth (resp.\ analytic) and periodic}?

\bigskip


The special case $m=1$ (\ie when $P(k)$ is the spectral projector corresponding to a
non-degenerate band $E_n$)  corresponds to an old problem in solid state physics,
namely the existence of smooth  and periodic Bloch functions.
Indeed, the solution of the eigenvalue problem
\begin{equation}\label{Eigenv equation}
    H(k) \psi_n(k) = E_n(k) \psi_n(k),
\end{equation}
yields a Bloch function $\psi_n(k)$ which is defined only up to a $k$ dependent
phase. Clearly one can always choose the phase in such a way that $\psi_n(k)$ is
\emph{locally} smooth in $k$, but it is not clear a priori if such
local solutions can be glued together to obtain a smooth and
periodic function. A geometrical obstruction might appear. For example, if one
includes a magnetic field in the Hamiltonian (thus breaking time-reversal symmetry)
it turns out that Question (A) has in general a negative answer, even in the smooth
case \cite{DuNo, No, Li}.

As for the time-symmetric Hamiltonian \eqref{Hamiltonian}, G.~Nenciu  proved that
the question has a positive answer, in the analytic sense, if $m=1$ or,
alternatively, $d=1$ (\cite{Ne83}, see also \cite{Ne91} Theorem 3.5 and references
therein). An alternative proof has been later provided by Helffer and Sj\"{o}strand
\cite{HeSj_Scroedinger}.


On the other side, in dimension $d=3$ the case of a non-degenerate Bloch band
globally isolated from the rest of the spectrum is not generic. It is more natural
to consider rather a family of Bloch bands which may cross each other, which means
to deal with  the case $m>1$.

In this paper we show that Question (A) has a positive answer in the analytic sense
for any $m \in \N$, provided that $d \leq 3$ and that the Hamiltonian satisfies
time-reversal symmetry. Borrowing the terminology introduced in
\cite{deCloizeaux64}, this can be rephrased by saying that we prove the existence of
analytic and periodic \emph{quasi-Bloch functions}.

The result is not a purely abstract one and has many applications.
Indeed the existence of analytic and periodic Bloch functions is the
premise for proving the existence of exponentially localized Wannier
functions \cite{Ne91, Thouless}. Similarly, a positive answer, in
the smooth sense, in the case $m >1$ is relevant for a rigorous
derivation of the semiclassical model of solid state physics
\cite{PST_3}, for the analysis of piezoelectricity in crystalline
solids \cite{PST_4}, and for the derivation of an effective
Hamiltonian for particles with spin degrees of freedom in a periodic
environment, \eg the Pauli equation or the Dirac equation with
periodic potential \cite{Ma}.

While Nenciu's proof exploits operator-theoretic techniques, our
strategy is to reformulate the problem in a bundle-theoretic
language, and then use Steenrod's classification theory
\cite{Steenrod} and some ideas in \cite{AvisIsham} in order to solve
it. It is our belief that mathematical physics benefits from the
interplay between analytic and geometric techniques, and we hope
that this result illustrates this viewpoint.

In Section 2 we state and prove our main results, which are then applied to the
specific case of Schr\"{o}dinger operators in Section 3 and to Dirac operators in Section 4.

\bigskip

\noindent \textbf{Acknowledgments.} I am gratefully indebted with H.\ Spohn and S.\
Teufel for suggesting to me to investigate this problem during the preparation of
\cite{PST_3}, and for many useful discussions. It is a pleasure to thank G.\
Dell'Antonio, B.\ Dubrovin, A.\ Ya.\ Maltsev  for stimulating discussion during the
very preliminary phase of this work. Last but not least, I am grateful to R.\
Percacci for a useful explanation about a result cited in his very recommendable
book.

\medskip

\noindent Financial support from DFG and HYKE-project is gratefully acknowledged.


\section{The main result}
\label{Sec main}

\medskip

\subsection{Assumptions and statements}

It is convenient to abstract from the specific context of
Schr\"{o}dinger-Bloch operators, and to state the result in a
general framework. Hereafter, we denote as $\B(\Hi)$ the algebra of
bounded operators over a separable Hilbert space $\Hi$, and with
$\U(\Hi)$ the group of unitary operators over $\Hi$. In the
application to Schr\"{o}dinger operators, one identifies $\Lambda$
with $\Gamma^{*}$.

\begin{assumption}[P] Let $\Lambda$ be a maximal lattice in $\R^d$. We assume
that $\{ P(k) \}_{k \in \R^d}$ is a family of orthogonal projectors
acting on a separable Hilbert space  $\Hi$, such that

\begin{description}
    \item{$\mathrm{(P_1)}$} the map $k \mapsto P(k)$ is smooth from $\R^d$ to $\B(\Hi)$
    \item{$\mathrm{(P_2)}$} the map $k \mapsto P(k)$ is covariant with respect to a unitary
    representation of the group $\Lambda$, in the sense that
    \[
    P(k + \la) = \ta(\la)^{-1} \, P(k) \, \ta(\la)  \qquad \forall k
    \in \R^d, \forall \la \in \Lambda,\]
    where $\ta: \Lambda \rightarrow  \U(\Hi)$ is a group
    homomorphism.  \hfill $\Box$
\end{description}
\end{assumption}


\goodbreak
\medskip

\noindent We are now in position to state our main result.

\begin{theorem} \label{Th main}
Let $\Lambda$ be a maximal lattice in $\R^d$. Let $\{ P(k) \}_{k \in \R^d}$ be a
family of orthogonal projectors acting on a separable Hilbert space  $\Hi$,
satisfying  Assumption {$\mathrm{(P)}$} and moreover:
\begin{description}
    \item[{$\mathrm{(P_3)}$}] there exists an antiunitary
    operator\footnote{\quad By \emph{antiunitary} operator we mean an antilinear operator $C: \Hi \to \Hi$,
     such that
    $\< C\ph, C \psi \>_{\Hi} = \< \psi, \ph \>_{\Hi}$ for any $\ph, \psi \in \Hi$.} 
     $C$ acting on $\Hi$ such that
    \[
    P(-k) =  C \, P(k) \, C  \qquad \mbox{ and  } \qquad C^{2} =1.
    \]
\end{description}

\noindent Let $m:= \dim P(k)$ and assume $d \leq 3, m \in \N$ or, alternatively, $d
\geq 4, m = 1$. Then each of the following equivalent properties holds true:

\renewcommand{\labelenumi}{{\rm(\Alph{enumi})}}
\begin{enumerate}
    \item \textbf{existence of global (quasi-)Bloch functions:} there exists a collection
    of smooth  maps $k \mapsto \ph_a(k)$ (indexed by  $a = 1, \ldots, m$) from $\R^d$ to $\Hi$  such that:
        \begin{description}
            \item {$\mathrm{(A_1)}$} the family $\{ \ph_a(k)\}_{a=1}^{m}$ is an orthonormal  basis spanning
            $\mathrm{Ran} P(k)$;
            \item {$\mathrm{(A_2)}$} each map is $\tau$-equivariant in the sense that
                   \[ \ph_a(k + \la) = \ta(\la)^{-1} \ph_a(k) \qquad \qquad \forall k
                   \in \R^d, \forall \la \in \Lambda. \]
        \end{description}
    \item \textbf{existence of an intertwining unitary:} there exists
    a smooth map $k \mapsto U(k)$ from $\R^d$ to $\U(\Hi)$ such that:
        \begin{description}
            \item {$\mathrm{(B_1)}$} each $U(k)$  intertwines $\Ran P(0)$ and
                  $\Ran P(k)$,
                    \[
                    U(k)^{*} \, P(k) \, U(k) = P(0)                       \qquad \forall k  \in \R^d;
                    \]
             \item{$\mathrm{(B_2)}$} the correspondence is  $\tau$-equivariant in the sense that:
                    \[
                     U(k + \la) = \ta(\la)^{-1} \, U(k)    \qquad \forall k \in \R^d, \forall \la \in \Lambda.
                    \]
        \end{description}

\end{enumerate} \hfill $\Box$
\end{theorem}
\goodbreak


It is convenient to reformulate properties $\mathrm{(A)}$ and
$\mathrm{(B)}$ in a bundle-theoretic language, by introducing the
complex vector bundle canonically associated to the family $\{ P(k)
\}_{k \in \R^d}$. More formally, for any family of projectors
satisfying Assumption $\mathrm{(P)}$, we define a hermitian complex
vector bundle $\vartheta$ in the following way. First one introduces
on the set $\R^d \times \Hf$ the equivalence relation $\sim_{\ta}$,
where
\[
(k,\ph) \sim_{\ta} (k', \ph')  \qquad  \Leftrightarrow  \qquad (k',
\ph')= (k + \la \,, \, \ta(\la) \ph) \quad \mbox{for some } \la \in
\Lambda.
\]
The equivalence class with representative $(k, \ph)$ is denoted as $[k,\ph]$. Then
the total space $E$ of the bundle $\vartheta$ is defined as
\[
E := \left \{ [k, \ph] \in (\R^d \times \Hf )/{\sim_{\ta}} : \quad
\ph \in \Ran P(k) \right \}.
\]
This definition does not depend on the representative in view of the covariance
property $\mathrm{(P_2)}$. The base space is the flat torus $\Base := \R^d /
\Lambda$ and the projection to the base space $\pi: E \to \Base $ is  $\pi[k, \ph] =
\mu(k)$, where $\mu$ is the projection modulo  $\Lambda$, $\mu: \R^d \to \Base$. One
checks that $\vartheta = (E \stackrel{\pi}{\rightarrow} \Base)$ is a smooth complex
vector bundle with typical fiber $\C^m$. In particular, the local triviality
follows, for example, from $\mathrm{(P_1)}$ and the use of the Nagy
formula\footnote{\quad Indeed, for any $k_0 \in \R^d$ there exist a neighbourhood
$\W \subset \R^d$ of $k_0$ such that $\| P(k) - P(k_0) \| < 1$ for any $k \in \W$.
Then by posing (Nagy's formula)
\[
W(k) := \( 1 - (P(k) - P(k_0))^2 \)^{-1/2} \left( P(k)P(k_0) + (1 - P(k))(1 -
P(K_0))\right)
\] one gets a smooth map $W: \W \rightarrow \U(\Hf)$ such that $W(k) \, P(k) \, W(k)^{-1} = P(k_0)$.
If $\{ \chi_{a} \}_{a=1,\dots,m}$ is any orthonormal basis spanning ${\mathrm
Ran}P(k_0)$,
then $\ph_a(k) = W(k) \chi_a$ is a smooth local orthonormal frame for $\vartheta$.}. 

 Moreover the vector bundle $\vartheta$ carries a natural hermitian
structure. Indeed, if $v_1, v_2 \in E$ are elements of the fiber over $x \in \Base$,
then up to a choice of the representatives
\[
v_1 = [x, \ph_1] \qquad \mbox{ and } \quad v_2 = [x, \ph_2],
\]
and one poses
\[
\< v_1, v_2 \>_{E_x} := \<  \ph_1, \ph_2 \>_{\Hi}.
\]

\noindent Endowed with this hermitian structure $\vartheta$ is turned into a
$G$-bundle with structural group $G=U(m)$.

\begin{proposition}
Under the same assumptions as in Theorem~\ref{Th main}, the properties
$\mathrm{(A)}$ and $\mathrm{(B)}$ are equivalent to:
\begin{description}
\item{$\mathrm{(C)}$} \textbf{triviality of the corresponding vector bundle:}
    the vector bundle associated to the family $\{ P(k) \}_{k \in \R^d}$ according to the previous construction
    is trivial in the category of smooth $U(m)$-bundles over $\Base$.
\end{description}
\hfill $\Box$
\end{proposition}

\begin{proof}
$\mathbf{(A)\Leftrightarrow (C).}$ Property (A) claims that the bundle $\vartheta$
admits a global smooth orthonormal frame, \ie that the principal bundle associated
to $\vartheta$ (\ie the \emph{bundle of frames} in the physics language) admits a
global smooth section. The latter claim is equivalent to the triviality of
$\vartheta$ in the category of smooth $U(m)$-bundles over $\Base$, namely property
(C).

\medskip

\noindent $\mathbf{(A)\Leftrightarrow (B).}$  Assume property (B). If $\{ \chi_a
\}_{a=1, \ldots ,m}$ is any orthonormal basis of $\Ran P(0)$, then $\ph_a(k) := U(k)
\chi_a$, for $a=1, \ldots,m$, satisfies condition (A). Viceversa, assume $\{ \ph_a
\}_a$ satisfies property (A). Then by posing
\[
W(k)\psi = \sum_a  \<\ph_a(0), \, \psi \>_{\Hi} \ph_a(k)
\]
one defines a partial isometry from $\Ran P(0)$ to $\Ran P(k)$. The orthogonal
projection $Q(k) := 1 - P(k)$ satisfies assumptions
$\mathrm{(P_1)}$-$\mathrm{(P_3)}$ too, since $C^{\, 2} = 1$. Thus, by the same
argument as before one gets a partial isometry $Y(k)$ intertwining $\Ran Q(0)$ and
$\Ran Q(k)$. By direct sum one gets a unitary operator $U(k) = W(k) \oplus Y(k)$
which satisfies property (B).

\end{proof}

\goodbreak

The proof of Theorem \ref{Th main} is based on the following scheme.
In the first part, by using standard ideas, one shows that
hypothesis $\mathrm{(P_3)}$ (which corresponds to
\emph{time-reversal symmetry} in the applications) implies that the
trace of the curvature of the Berry connection of $\vartheta$ has a
special property, namely $\Omega(-k) = -\Omega(k)$. Thus the first
Chern class of $\vartheta$ vanishes. The difficult step is to show
that this condition is \emph{sufficient} for the triviality of the
bundle $\vartheta$. The latter claim, whose proof is based on
Proposition \ref{Prop Abstract classification}, relies on the
special structure and the low-dimensionality of the base space $B
\approx \T^d, d \leq 3$. (In this paper the symbol $\approx$ denotes
homeomorphism of topological spaces)

\goodbreak \medskip

By the \emph{Oka's principle}, the result can be pushed forward to
the analytic category, yielding the following ``corollary".

\begin{theorem} \label{Th analytic}
Let $\mathcal{T}_a = \{ z \in \C^d: |\, \mathrm{Im} z_i \,| <a, \, \forall i
=1,\ldots,d \}$ for a fixed $a >0$ and $\Lambda$ a maximal lattice in $\R^d$,
regarded as a subset of $\C^d$. Let $\{ P(z)\}_{z \in \t_a}$ be a family of
orthogonal projectors in $\Hi$, satisfying
\begin{description}
    \item{$\mathrm{(\widetilde{P}_1)}$} the map $z \mapsto P(z)$ is \emph{analytic} from $\t_a$ to $\B(\Hi);$
    \item{$\mathrm{(\widetilde{P}_2)}$} the map $z \mapsto P(z)$ is $\tau$-covariant, in the sense that
    \[
    P(z + \la) = \ta(\la)^{-1} \, P(z) \, \ta(\la)  \qquad \forall z
    \in \t_a, \forall \la \in \Lambda,\]
    where $\ta: \Lambda \rightarrow  \U(\Hi)$ is a group
    homomorphism;
    \item{$\mathrm{(\widetilde{P}_3)}$} there exists an antiunitary operator $C$ acting on $\Hi$ such that
    $C^{2} =1$ and $P(-z) =  C \, P(z) \, C$ for all $z \in \t_a$.
\end{description}
\noindent Let $m:= \dim P(z)$ and assume $d \leq 3, m \in \N$ or, alternatively, $d
\geq 4, m = 1$. Then each of the following equivalent properties holds true:
\renewcommand{\labelenumi}{{\rm(\Alph{enumi})}}
\begin{enumerate}
    \item there exists a collection
    of \emph{analytic} functions $z \mapsto \ph_a(z)$ (indexed by  $a = 1, \ldots, m$) from $\t_a$ to $\Hi$
    satisfying $\mathrm{(A_1)}$ and $\mathrm{(A_2)}$ over $\t_a$;
    \item there exists an \emph{analytic} function $z \mapsto U(z)$ from $\t_a$ to $\U(\Hi)$ satisfying
    $\mathrm{(B_1)}$ and $\mathrm{(B_2)}$ over $\t_a$.
\end{enumerate}  \hfill $\Box$
\end{theorem}

\noindent Notice that Theorem~\ref{Th analytic} provides a complete answer, for $d
\leq 3$, to the question raised in \cite{Ne83}.

\goodbreak


\subsection{Proof of main results}

\begin{proof}[Proof of Theorem 1]
Let $\Omega$ be the differential $2$-form over $\R^d$ with
components
\[
\Omega_{i,j}(k) =  \Tr \(P(k) \, [\d_i P(k), \, \d_j P(k)] \)
\] \ie
\be  \label{Omega} \Omega(k) = \sum_{i,j} \Omega_{i,j}(k)\, dk^i \wedge dk^j. \ee In
view of property $\mathrm{(P_2)}$,  $\Omega$ is $\Lambda$-periodic, and thus defines
a 2-form over $\Base$. We are going to show how $\Omega$ is related to the curvature
of a connection over the vector bundle $\vartheta$.

\medskip


By  using a local frame $\Psi = (\psi_1, \ldots, \psi_m)$  over $\W \subset \R^d$,
 one defines locally a 1-form $\A(k) = \sum_{i} \A_i(k) dk^i$ with
coefficients $\A_i(k)$ in $\frak u(m)$, the Lie algebra of
antihermitian matrixes, given by\footnote{\quad Here and in the
following $i,j, \ldots \indexd$ are the base-space indexes, while
$a,b,c \in \{1, \ldots, m \}$ are the matrix (Lie algebra) indexes.}
\be \A_i(k)_{ab} =  \< \psi_a(k), \d_i \psi_b(k)\>, \qquad k \in \W.
\ee It is easy to check how $\A$ transforms under a change of local
trivialization: if $\widetilde{\Psi}=(\tilde{\psi}_1, \ldots,
\tilde{\psi}_m)$ is a local trivialization over $\tilde{\W}$, such
that $\Psi(k) = G(k) \tilde{\Psi}(k)$ for a smooth $G:\W \cup
\widetilde{\W} \to U(m)$, then the 1-form $\A$ transforms as \be
\label{A transforms} \widetilde{\A_i}(k) =  G(k)^{-1} \, \A_i(k) \,
G(k) + G(k)^{-1}\, dG(k)   \qquad k \in \W \cap \widetilde{\W}. \ee
The transformation property (\ref{A transforms}) implies (see
\cite{Bleecker}, Theorem 1.2.5) that $\A$ is the local expression of
a $U(m)$-connection over the complex vector bundle $\vartheta$.
(Such a connection is called \emph{Berry connection} in the physics
literature. Mathematically, it is the connection induced by the
embedding of $\vartheta$ in the trivial hermitian bundle $\Base
\times \Hf \to \Base$).

\goodbreak

A lengthy  but straightforward computation yields
\[
\begin{array}{ccc}
  \Omega_{i,\,j}
                 & = & \tr \( \d_i \A_j - \d_j \A_i + \A_i \, \A_j - \A_j \, \A_i \)
\end{array}
\]
where $\tr$ denotes the trace over the matrix (Lie algebra) indexes. Thus one concludes
that $\Omega = \tr \omega_{\A}$, where
\[
\omega_{\A} := d\A + \A \wedge \A
\]
represents locally the curvature of the connection $\A$.
Therefore the first real Chern class of the bundle $\vartheta$ is
\[
\mathrm{Ch}_1(\vartheta) = \footnotesize{\frac{i}{2\pi}} \,\, [\tr \omega_{\A}] =  \frac{i}{2\pi} \,\, [\Omega],
\] where $[\ldots]$ denotes the de Rahm cohomology class.


\smallskip

By property $\mathrm{(P_3)}$ one has that $\d_iP(-k) = - C \, \d_i
P(k) \, C$, thus
\[
\begin{array}{lll}
\Omega_{i,j}(-k) & = &  \Tr \( C \, P(k) C \, C \, [\d_i P(k), \, \d_jP(k)]\, C  \) \\ [3 mm]
                 & = &  - \Tr \( P(k) \, [\d_i P(k), \, \d_j P(k)] \) \\ [3 mm]
                 & = &  - \Omega_{i,j}(k),
\end{array} \] where we used  the fact that $ \Tr (C\, A \, C) = \Tr (A^*)$
for any $A \in \B(\Hi)$. Thus one concludes that \be \label{Omega
symmetry} \Omega(-k) = - \Omega(k). \ee


\medskip \goodbreak

It follows from (\ref{Omega symmetry}) that the first \emph{real}
Chern class of $\vartheta$ vanish. Indeed, in $B \approx \T^d$
equipped with periodic coordinates $k=(k_1,\ldots, k_d), \, k_i \in
[-\pi,\pi)$, one considers the 2-cycles defined by the sets \be
\label{Cycles define} \Theta_{j,l} := \left\{k \in \T^d : k_i=0
\mbox{ for any $i \notin \{j,l\}$} \right\}, \qquad \mbox{for } j,l=
1,\ldots,d, \, j \neq l, \ee with any consistent choice of the
orientation. From (\ref{Omega symmetry}) it follows that \be
\label{Omega vanishing} \frac{i}{2 \pi} \int_{\Theta_{j,l}} \Omega =
0. \ee It remains to show that the independent cycles $\{
\Theta_{j,l} \}_{i \neq j}$ are a basis for $H_2(\T^d, \R)$. Indeed,
from K\"{u}nneth formula one proves by induction that $H_2(\T^d, \Z)
\cong \Z^{k(d)}$ with $k(d)= {\frac{1}{2}d(d-1)}$. Therefore, the
independent 2-cycles $\Theta_{j,l}$ generate, by linear combinations
with coefficients in $\Z$ (resp.\ $\R$), all $H_2(\T^d, \Z)$ (resp.
$H_2(\T^d, \R)$). Thus, by de Rham's isomorphism theorem, from
(\ref{Omega vanishing}) it follows that $\mathrm{Ch}_1(\vartheta)=0$

\goodbreak
\medskip


We conclude that the first \emph{real} Chern class of the bundle
$\vartheta$ vanishes. Since the natural homomorphism $H^2(\T^d, \Z)
\to H^2(\T^d, \R)$ is injective, this implies the vanishing of the
first \emph{integer} Chern class.

As for $m=1$, it is a classical result by Weil and Constant
(\cite{Weil}, see also \cite{Brylinski} Theorem 2.1.3) that the
vanishing of the first integer Chern class of a complex line bundle
over a (paracompact) manifold implies the triviality of the bundle
itself. For $m \geq 2$, it follows from Proposition \ref{Prop
Abstract classification} that for a base space $B \approx \T^d$ with
$d \leq 3$ the vanishing of the first real Chern class implies the
triviality of the bundle $\vartheta$, \ie property (C). This
concludes the proof of the Theorem.
\end{proof}


\bigskip \goodbreak

\begin{proof}[Proof of Theorem 2] In strict analogy with the smooth case, the
problem is equivalent to the triviality (in the analytic category)
of an analytic $U(m)$-bundle $\tilde{\vartheta}$ over the open
poly-cylinder $\t_a/\Lambda$. The proof of Theorem~1 implies that
$\tilde{\vartheta}$ is trivial in the category of \emph{smooth}
$U(m)$-bundles over $\t_a/\Lambda$.

By the Oka principle (see \cite{FritzscheGrauert}, Chapter V) if an
analytic bundle over a Stein manifold is topologically trivial, then
it is analytically trivial. This result applies to our case, since
$\t_a/ \Lambda$ is the cartesian product of non-compact Riemann
surfaces, and as such a Stein manifold.
\end{proof}


\subsection{A technical lemma}

We prove in this section a technical result used in the proof of Theorem \ref{Th
main}, which shows that when the base space is a a low dimensional torus (or, more
generally, any low dimensional connected compact manifold whose second cohomology is
torsionless) the vanishing of the first real Chern class of a $U(m)$-bundle implies
the triviality of the bundle itself. The proof is based on Steenrod's classification
theory \cite{Steenrod} and on some ideas in the literature \cite{AvisIsham}.

\medskip

We first recall (\cite{Spanier} Section 5.9) that there is a natural transformation
$i: H^2(\, \cdot \,, \Z) \stackrel{}{\rightarrow} H^2(\, \cdot \,, \R)$, so that for
any $f: X \stackrel{}{\longrightarrow}  Y$ the following diagram is commutative:
\begin{align*}
\bfig
\node h2xz(0,0)[H^2(X,\Z)]
\node h2xr(0,-400)[H^2(X,\R)]
\node h2yz(750,0)[H^2(Y,\Z)]
\node h2yr(750,-400)[H^2(Y,\R)]
\arrow[h2yz`h2xz;f^{\ast}]
\arrow[h2yr`h2xr;f^{\ast}]
\arrow[h2xz`h2xr;i_X]
\arrow[h2yz`h2yr;i_Y]
\efig
\end{align*}
When one specialize to $X \cong \T^d$, the natural homomorphism
$i: H^2(\T^d, \Z) \rightarrow H^2(\T^d,\R)$ is injective. \label{Lemma injectivity}

We denote as $k_G(X)$ the set of vertical isomorphism classes of principal smooth
$G$-bundles over $X$ (see \cite{Husemoller}, Section 4.10).  By  vertical
isomorphism we mean an isomorphism which projects over the identity map on $X$, \ie
reshuffling of the fibers is not allowed.
\smallskip

\begin{proposition} \label{Prop Abstract classification}
If $X$ is a compact, connected manifold of dimension $d \leq 3$ and $G = U(m)$ for
$m \geq 2$, then $k_G(X) \cong H^2(X,\Z)$, where the isomorphism (of pointed sets)
is realized by first integer Chern class. In particular, if  $X$ is such that the
natural homomorphism $H^2(X,\Z) \rightarrow H^2(X,\R)$ is injective, then for any
$U(m)$-bundle $\vartheta$ over $X$ the vanishing of the first \emph{real} Chern
class $\mathrm{Ch}_1(\vartheta)$ implies the triviality of $\vartheta$.
\hfill $\Box$
\end{proposition}

For sake of a more readable proof, we first recall some results about the
classification theory of $G$-bundles \cite{Steenrod}. A principal $G$-bundle $
\Upsilon_G =(E_G \stackrel{p_G}{\rightarrow} B_G; G)$ is said to be \emph{universal}
if the map $[X, B_G] \rightarrow k_G(X)$, which associate to a (free) homotopy class
of maps $[f]$ the isomorphism class of the pull-back bundle $f^{*}\Upsilon$, is a
bijection for all $X$. A principal $G$-bundle with total space $P$ is universal if
and only if $P$ is contractible, and for any finite-dimensional Lie group $G$ there
exists a universal $G$-bundle. The base spaces of different universal $G$-bundles
for the same group $G$ are homotopically equivalent.

\smallskip

We also make use in the proof of the Eilenberg-Mac Lane spaces
(see \cite{Spanier}, Sec. 8.1). We recall that for any $n \in \N$
and any group $\pi$ (abelian if $n \geq 2$) there exists a path
connected space $Y$ such that $\pi_k(Y)= \pi$ for $k=n$ and zero
otherwise. This space is unique in the category of CW-complexes
and denoted by $K(\pi,n)$.

\begin{proof} From abstract classification theory we know that $k_G(X) \cong
[X,B_G]$, but unfortunately a simple representation of $[X, B_G]$
is generally not available. The crucial observation
\cite{AvisIsham} is that if we are interested only in manifolds
with $\dim X \leq n$ the homotopy groups of $B_G$ beyond the
$n^{\rm th}$ do not play any role, therefore one can "approximate"
$B_G$ with a space $B_3$ which captures the relevant topological
features of $B_G$.

More precisely, one constructs a space $B_3$ which is $4$-equivalent to $B_G$, in
the sense that there exist a continuous map \[ \rho: B_G \longrightarrow B_3 \] such
that \[ \pi_k(\rho): \pi_k(B_G) \longrightarrow \pi_k(B_3)  \] is an isomorphism for
$k \leq 3$ and a epimorphism  for $k=4$. Therefore, for any complex $X$ of dimension
$d \leq 3$, one has $[X,B_G]=[X,B_3]$.

\smallskip

From the exact homotopy sequence of the universal bundle $\Upsilon$ one has
$\pi_k(B_G)= \pi_{k-1}(G)$, so that for $G=U(m)$ one has
\begin{enumerate}
  \item $\pi_1(B_G)= \pi_{0}(G)=0$,
  \item $\pi_2(B_G)= \pi_{1}(G)= \Z,$
  \item $\pi_3(B_G)= \pi_{2}(G)= 0. $
\end{enumerate}

Since $B_G$ is simply connected, there is already a
$2$-equivalence
\[  \rho: B_G \longrightarrow B_3:=K(\Z,2)\thickapprox \C P^{\infty} \]
see \cite{Spanier}. Since $\pi_3(B_G)=0$, $\pi_3(\rho)$ is an
isomorphism, and $\pi_4(\rho)$ is surjective since
$\pi_4(K(\Z,2))=0$. Therefore $\rho$ is a $4$-equivalence,  so
that \[ k_G(X) \cong [X, K(2,\Z)] \cong H^2(X,\Z). \]

The first identification is an isomorphism of pointed sets, \ie the trivial element
$[f] \in [X, K(2,\Z)]$ corresponds to the (equivalence class of) the trivial
$U(m)$-bundle over $X$. As for the second, let be $\eta$ any non zero element of
$H^2(\C P^{\infty}, \Z)$. Then, according to \cite{Spanier} Theorem 8.1.8, the map
\[
\begin{array}{cccc}
  \psi_X: & [X,K(2,\Z)] & \rightarrow & H^2(X, \Z) \\
    & [f] & \mapsto & f^{*} \eta  \\
\end{array}
\] is an isomorphism of pointed sets. Consider now the following diagram
\begin{align*}
\bfig
\node xk2z(0,0)[{[X,K(2,\Z)]}]
\node h2xz(850,0)[H^2(X,\Z)]
\node kGX(0,-500)[k_G(X)]
\node h2xr(850,-500)[H^2(X,\R)]
\arrow[xk2z`h2xz;\psi_X]
\arrow[kGX`h2xr;\mathrm{Ch}_1]
\arrow[xk2z`kGX;]
\arrow[h2xz`h2xr;i_X]
\arrow[kGX`h2xz;\mathrm{ch}_1]
\efig
\end{align*}
where the diagonal arrow represents the first integer Chern
class. The lower tringle is
commutative since $\mathrm{Ch}_1 = i \circ \mathrm{ch}_1$. As for the upper
triangle, one choose $\eta := \mathrm{ch}_1(\Upsilon_G)$ which is certainly not
zero. Then, since
\[  \mathrm{ch}_1(f^{*}\Upsilon_G) = f^{*}\mathrm{ch}_1(\Upsilon_G)=f^{*}\eta,
\] the upper triangle is commutative. Thus $\mathrm{ch}_1$ is an isomorphism of pointed sets.

Finally, if $\mathrm{Ch}_1(\vartheta)=0$ then the injectivity of $i_{\T^d}$ implies
that $\mathrm{ch}_1(\vartheta)=0$. Since $\mathrm{ch}_1$ is an isomorphism of
pointed sets, $\vartheta$ must be the distinguished point in $k_G(X)$, namely the
isomorphism class of the trivial $U(m)$-bundle over $X$.
\end{proof}

\goodbreak
\newpage


\section{Application to Schr\"{o}dinger operators}

In this Section, we comment on the application of the general results to
Schr\"{o}dinger operators in the form \eqref{Hamiltonian}. The lattice $\Gamma$ is
represented as
\[
\Gamma =\Big\{ x\in\R^d: x= \textstyle{\sum_{j=1}^d}\alpha_j\,\gamma_j
\,\,\,\mbox{for some}\,\,\alpha \in \mathbb{Z}^d \Big\}\,,
\] where $\{\gamma_1,\ldots,\gamma_d \}$ are independent vectors in $\R^d$.
We denote by $\Gamma^*$ the dual latice of $\Gamma$ with respect to the standard
inner product in $\R^d$, \ie the lattice generated by the dual basis
$\{\gamma_1^*,\ldots,\gamma_d^*\}$ determined through the conditions $\gamma_j^*
\cdot \gamma_i = 2\pi \delta_{ij}$, $i,j\indexd$. The centered fundamental domain
$Y$ of $\Gamma$ is defined by
\[
 Y = \Big\{ x\in\R^d: x=
\textstyle{\sum_{j=1}^d}\beta_j\,\gamma_j
\,\,\,\mbox{for}\,\,\beta_j\in
[-\textstyle{\frac{1}{2},\frac{1}{2}}]
 \Big\}\,,
\]
 and analogously the centered fundamental domain $Y^{*}$ of $\Gamma^*$.  The set $Y^*$ is usually called
 the {\em first Brillouin zone} in the physics literature.

\medskip

\subsection{The Bloch-Floquet-Zak representation}

As usual in the  recent mathematical literature, we use a variant of
the Bloch-Floquet transform, which is called the Bloch-Floquet-Zak
transform, or just the Zak transform for sake of brevity. The
advantage of such a variant is that the fiber at $k$ of the
transformed Hamiltonian operator has a domain which does not depend
on $k$.

\noindent The Bloch-Floquet-Zak transform is defined as \be
\label{Zak transform} (\UZ\psi)(k,x):=\sum_{\gamma\in\Gamma} \E^{-\I
k \cdot (x+\gamma)} \, \psi(x+\gamma), \qquad (k,x)\in\R^{2d}, \ee
initially for a fast-decreasing function $\psi\in\Sch(\R^d)$. One
directly reads off from (\ref{Zak transform}) the following
periodicity properties \be \label{Zak1} \big(\UZ\psi\big) (k,
y+\gamma) = \big( \UZ\psi\big) (k,y)\quad \mbox{ for all} \quad
\gamma\in\Gamma\,, \ee \be \big(\UZ\psi\big) (k+\lambda, y) =
\E^{-\I y\cdot\lambda}\,\big( \UZ\psi\big) (k,y) \quad\mbox{ for
all} \quad \lambda\in\Gamma^*\,. \label{Zak2} \ee

\noindent From (\ref{Zak1}) it follows that, for any fixed $k\in{\R^d}$, $\big(
\UZ\psi \big)(k,\cdot)$ is a $\Gamma$-periodic function and can thus be regarded as
an element of $\Hf := L^2(T_Y)$, $T_Y$ being the flat torus $\R^d/\Gamma \approx
\T^d$.

\noindent On the other side, equation (\ref{Zak2}) involves a unitary representation
of the group of lattice translations on $\Gamma^*$ (isomorphic to $\Gamma^*$ and
denoted as $\Lambda$), given by
\begin{equation} \label{tau definition}
\tau:\Lambda \to \U(\Hf)\,,\quad\lambda \mapsto \tau(\lambda)\,, \quad
\big(\tau(\lambda)\ph \big)(y) = \E^{\I\,y\cdot\lambda} \ph(y).
\end{equation}
It is then convenient to introduce the Hilbert space

\be   \label {H tau} \Hi_\tau :=\Big\{ \psi\in L^2_{\rm loc}(\R^d, \Hf ):\,\, \psi(k
- \la) = \tau(\la)\,\psi(k) \qquad \forall \la \in \Lambda \Big\}\,,  \ee equipped
with the inner product
\[
\langle \psi,\,\ph\rangle_{\Hi_\tau} = \int_{Y^{*}} dk\, \langle
\psi(k),\,\ph(k)\rangle_{\Hf}\,.
\]
Obviously, there is a natural isomorphism  between $\Hi_\tau$ and $L^2(Y^{*}, \Hf)$
given by restriction from $\R^d$ to $Y^{*}$, and with inverse given by
$\tau$-equivariant continuation, as suggested by (\ref{Zak2}).  Equipped with these
definitions, one checks that the map defined by (\ref{Zak transform}) extends to a
unitary operator
\[
\UZ: L^2(\R^d)\to \Ht \cong L^2(Y^{*}, L^2(T_Y)),
\]
with inverse given by
\[
\( \UZ^{-1} \ph \)(x) = \int_{Y^*} dk \, \E^{\I k \cdot x} \ph(k, [x]),
\]
where $[\, \cdot \, ]$ refers to  the a.e. unique decomposition $x = \gamma_x + [x]$,
with $\gamma_x \in \Gamma$ and $[x] \in Y$.

As  mentioned in the introduction, the advantage of this construction is that the
transformed Hamiltonian is a fibered operator over $Y^*$.  Indeed, for the  Zak
transform of the  Hamiltonian operator (\ref{Hamiltonian}) one finds
\[
\UZ \, H  \,  \UZ^{-1} = \int_{Y^{*}}^\oplus dk\,H_{\rm per}(k)
\]
with fiber operator \be \label{H(k)} H_{\rm per}(k) = {\footnotesize \frac{1}{2}}
\big( -\I \nabla_y + k\big)^2 + V_\Gamma(y)\,,  \quad k\in Y^{*} \,. \ee For fixed
$k\in Y^{*}$ the operator $H_{\rm per}(k)$ acts on $L^2(T_Y)$ with
domain\footnote{\quad We denote as $W^{k,p}(X)$ the Sobolev space consisting of
distributions whose $k$-th derivative is (identifiable  with) an element of
$L^p(X)$.} $W^{2,2}(T_Y)$ independent of $k\in Y^{*}$, whenever the potential
$V_{\Gamma}$ is infinitesimally bounded with respect to $-\Lap$. Under the same
assumption on $V_{\Gamma}$, each fiber operator $H(k)$ has pure point spectrum
accumulating at infinity: $E_0(k) \leq E_1(k) \leq E_2(k) \leq \ldots$

\bigskip

We denote as $\sigma_0(k)$ the set $\{ E_{i}(k): n \leq i \leq n+m-1 \}$,
corresponding to a physically relevant family of Bloch bands, and we assume the
following \emph{gap condition}:
\begin{equation}\label{Gap condition}
    \mathrm{dist}\( \sigma_0(k), \sigma(H(k)) \setminus \sigma_0(k) \) \geq g >0, \qquad
    \forall k \in Y^*.
\end{equation}

\bigskip

\noindent Let $ P(k) \in \B(\Hf)$ be the spectral projector of $H(k)$ corresponding
to the set $\sigma_0(k) \subset \R$. The family $\{ P(k) \}_{k \in \R^d}$ satisfies
assumption $\mathrm{(P_1)}$-$\mathrm{(P_3)}$ stated in Section~\ref{Sec main}.
Indeed, the map $k \mapsto P(k)$ is smooth from $\R^d$ to $\B(\Hf)$, since $H(k)$
depends smoothly (in the norm-resolvent sense) upon $k$, and the gap condition
\eqref{Gap condition} holds true. Moreover, from (\ref{H(k)}) one checks that
\[  \label{H covariant}
H(k + \la) = \ta(\la)^{-1} \, H(k)  \, \ta(\la), \qquad \forall \la \in \Lambda,
\]  and since $\sigma_0$ is periodic one concludes that
\be \label{P covariant} P(k + \la) = \ta(\la)^{-1} \, P(k)  \, \ta(\la), \qquad
\forall \la \in \Lambda. \ee


Property $\mathrm{(P_3)}$ corresponds to time-reversal symmetry. This symmetry
 is realized in $L^2(\R^d)$ by the complex conjugation operator, \ie by the operator
\[
( T \psi )(x) = \bar{\psi}(x), \qquad  \qquad \psi \in L^2(\R^d).
\]
By the Zak transform we get that $\tilde{T}= \UZ T \UZ^{-1}$ acts as
\[
( \tilde{T} \ph )(k) = C \, \ph(-k), \qquad  \qquad \ph \in L^2(Y^{*},\Hf),
\]
where $C$ is the complex conjugation operator in $\Hf$. Operators in the form
(\ref{Hamiltonian}) commute with the time-reversal operator $T$. The following
statement is analogous to a result proved in \cite{PST_4}. We repeat the proof for
the sake of completeness.

\begin{proposition}[\textbf{Time-reversal symmetry}] \label{Prop timerev}
Assume that the self-adjoint operator $H$ commutes with $T$ in $L^2(\R^d)$, and that
$\UZ H \UZ^{-1}$ is a \emph{continuously} fibered operator. Let $P(k)$ be the
eigenprojector of $H(k)$ corresponding to a set $\sigma_0(k)$, satisfying (\ref{Gap
condition}). Then \be \label{Timerev projector} P(k) = C \, P(-k) \, C. \ee
\end{proposition}

\begin{proof} The transformed Hamiltonian $\UZ H\UZ^{-1}$ commutes with $\tilde{T}$, yielding a symmetry of the fibers,
\ie \be \label{Timerev hamiltonian} H(k) = C H(-k)C. \ee
By definition, for any
Bloch band $E_i$ one has
\[
H(k) \ph(k) = E_i(k) \ph(k)
\]
for a suitable $\ph(k) \in \Hf, \, \ph(k) \neq 0$.  By complex conjugation one gets
\[
E_i(k) \, C \ph(k) = C H(k) \ph(k) = C H(k) C \,  C \ph(k) = H(-k) \, C \ph(k),
\] which shows that $E_i(k)$ is an eigenvalue of $H(-k)$. By the \emph{continuity} of
  $k \mapsto E(k,t)$ and the \emph{gap condition}, by
starting from $k=0$ one concludes that $E_i(-k) = E_i(k)$ for any $k$. Thus
$\sigma_0(-k) = \sigma_0(k)$.

Since $P(k) = \chi_{\sigma_0(k)} \( H(k) \)$,  where $\chi_{\sigma_0(k)}$ is a
smoothed characteristic function whose support contains $\sigma_0(k)$ and no other
point of the spectrum of $H(k)$, from (\ref{Timerev hamiltonian}) one gets
(\ref{Timerev projector}) by applying the functional calculus and noticing that $f(C
\, A \, C) = C \, f(A) \,C$ whenever $A$ is self-adjoint and $f$ is an admissible
function. \end{proof}

We conclude that, in the Zak representation, the family of projectors $\{ P(k)\}_{k
\in \R^d}$ corresponding to a relevant family of Bloch bands, satisfy assumptions
$\mathrm{(P_1)}$-$\mathrm{(P_3)}$ of Section~\ref{Sec main}.

\bigskip \goodbreak


\subsection{Comparison with the usual Bloch-Floquet formalism}

While from a mathematical viewpoint it is  convenient to use the
Bloch-Floquet-Zak  transform, as defined in (\ref{Zak transform}),
in the solid state physics literature one mostly encounters the
classical Bloch-Floquet transform, defined by
\begin{equation}\label{BF transform}
    (\UB \psi)(k,y):=\sum_{\gamma\in\Gamma} e^{- i
k \cdot \gamma }\psi(y+\gamma), \qquad (k,y)\in\R^{2d}
\end{equation}
initially for $\psi\in\Sch(\R^d)$. We devote this short subsection
to a comparison of the two choices.

Functions in the range of $\UB$ are periodic in $k$ and
quasi-periodic in $y$, \be \label{BF12} \big(\UB\psi\big) (k,
y+\gamma) = \E^{\I k\cdot\gamma}\,\big( \UB\psi\big) (k,y)\quad
\mbox{ for all} \quad \gamma\in\Gamma\,, \ee \be \big(\UB\psi\big)
(k+\lambda, y) = \big( \UB\psi\big) (k,y) \quad\mbox{ for all} \quad
\lambda \in\Gamma^*\,. \label{BF22} \ee

\noindent Definition (\ref{BF transform}) extends to a unitary operator \be \UB:
L^2(\R^d) \rightarrow \Hi_{\rm B} := \int_{Y^*}^{\oplus} \Hi_{k} \, dk \ee where \be
\Hi_{k} := \big\{ \ph \in L^2_{\rm loc}(\R^d): \ph(y+\gamma)= e^{i k \cdot \gamma}
\ph(y) \quad \forall \gamma \in \Gamma \big\}. \ee

\noindent Although we use the standard (but somehow misleading) ``direct integral"
notation, it is convenient to keep in mind that the space appearing on the
right-hand side is the Hilbert space consisting of the locally-$L^2$ sections of an
Hilbert space bundle with base space $Y^{*}$ (identified with a $d$-dimensional
torus) and whose fiber at point $k$ is $\Hi_k$.

\medskip

The relation between the Bloch-Floquet and the Zak representation is easily obtained
by computing the unitary operator $\J = \UB \UZ^{-1}$, which is explicitely given by
\[
\( \J \ph \)(k,y) = \E^{i k \cdot y} \ph(k,y).
\]

\noindent Clearly $\J$ is a fibered operator, whose fiber is denoted as $J(k)$.
Notice that $J(k)^{-1}$ maps unitarily the space $\Hi_k$ into the typical fiber
space $\Hi_0 = \Hf = L^2(T_Y)$. If $H_{\textrm{B}}(k)$ is the fiber of the
Hamiltonian $H$ in Bloch-Floquet representation, 
one has
\[
J(k) \, H_{\textrm{B}}(k) \, J(k)^{-1} = H_{\rm per}(k),
\]
see (\ref{H(k)}), and thus $\sigma \(H_{\rm per}(k) \) = \sigma \( H_{\rm B}(k) \)$.

\medskip \goodbreak

As for  the relevant family of projectors, we notice that an operator-valued
function $k \mapsto P_{\rm B}(k)$, with $P_{\rm B}(k) \in \B(\Hi_k)$, is periodic if
and only if $P_{\rm Z}(k) := J(k) \, P_{\rm B}(k) \, J(k)^{-1}$ is
$\tau$-equivariant with respect to the representation in  (\ref{tau definition}).
Moreover, conjugation with $\J$ (resp. with $\J^{-1}$) preserves smoothness and
analyticity, since $\J$ acts as a multiplication times a unitary operator $J(k)$
which depends analytically on $k$. Thus a family of orthogonal projectors $P_{\rm
B}(k)$ is smooth (resp.\ analytic) and periodic if and only if the corresponding
family $P_{\rm Z}(k)$ is smooth (resp.\ analytic) and $\tau$-covariant. The
results in Section~\ref{Sec main} thus directly apply to this situation, yielding
the existence of a smooth and periodic orthonormal basis for $\Ran P_{\rm B}(k)$.


\newpage

\section{Application to Dirac operators}

There are experiments in atomic and solid state physics where the relativistic
corrections to the dynamics of the electrons are relevant, while the energy scale at
which the experiment is performed is not so high to require the use of a fully
relativistic theory, namely Quantum Electrodynamics. Such physical situations are
conveniently described by using a hybrid model, which embodies some relativistic
effects (as, for example, the spin-orbit coupling) without involving the
difficulties of a fully relativistic theory.

In order to introduce the model, one first fixes an inertial frame, \eg the
laboratory frame. In such a frame, the potential to which the electron is subject is
described by the function $V: \R^d \to \R$. Then it is postulated that the dynamics
of the electron in the chosen frame is described by the Dirac equation
\[
i \psi_t = H_{\rm D} \,\, \psi_t, \qquad \qquad \psi_t \in L^2(\R^3, \C^4),
\]
with
\begin{equation}\label{Dirac Hamiltonian}
    H_{\rm D} = -ic \nabla \cdot \alpha + m_{\rm e} c^2 \, \beta + V,
\end{equation}
where $m_{\rm e}$ denotes the mass of the electron and $c$ the speed of light, and
where $\alpha =(\alpha_1, \alpha_2, \alpha_3)$ and $\beta$ are given by
\[
\alpha_i = \(\begin{array}{cc}
               0        & \sigma_i \\
               \sigma_i &   0
             \end{array}
\), \qquad \beta = \( \begin{array}{cc}
                        1_{\C^2} & 0 \\
                        0 & -1_{\C^2}
                      \end{array}\),
\]
with $(\sigma_1, \sigma_2, \sigma_3)$ the vector of Pauli spin matrixes.

Such a model  is clearly not Lorentz covariant, but it is expected to include the
relativistic  corrections of lowest order (in the parameter $c^{-1}$, as $c \to
\infty$) to the dynamics described by the Schr\"{o}dinger equation \cite{Thaller}.

\medskip

We now specialize to the case $V = V_{\Gamma}$, with $V_{\Gamma}$ periodic with
respect to a lattice $\Gamma \subset \R^3$. We set $m_{\rm e}=1$  and $c=1$ for
simplicity. As in the case of Schr\"{o}dinger operators, one introduces the
Bloch-Floquet-Zak transform, defined as in (\ref{Zak transform}), which yields a
unitary operator
\[
\UZ : L^2(\R^3, \C^4) \to L^2(Y^*, \Hf),
\]
where $\Hf = L^2(T_Y) \otimes \C^4$ with $T_Y := \R^3/\Gamma$.

\noindent The transformed Hamiltonian operator $\UZ H_{\rm D} \UZ^{-1}$ is fibered,
with fiber
\[
H_{\rm D}(k) = (-i \nabla + k) \cdot \alpha +  \beta + V_{\Gamma},
\]
acting in $\Hf$, with domain $H^1(T_Y, \C^4)$.  Under general assumptions on the the
periodic potential (\eg if $V_{\Gamma}$ is infinitesimally bounded with respect to
$i \nabla $), each fiber $H_{\rm D}(k)$ has compact resolvent and thus pure point
spectrum accumulating to infinity.  Since $H_{\rm D}(k)$ is not bounded from below,
the labelling of eigenvalues requires some additional care: one can prove that there
is a consistent global labelling $\{ \mathcal{E}_{n}(k) \}_{n \in \Z}$ such that
each $k \mapsto \mathcal{E}_n(k)$ is continuous and periodic, and the relation
$\mathcal{E}_n(k) \leq \mathcal{E}_{n+1}(k)$ holds true. We say that the function
$\mathcal{E}_n$ is the $n$-th \emph{Bloch-Dirac band}.

\medskip

Whenever the potential is reflection-symmetric, \ie $V_{\Gamma}(-x) =
V_{\Gamma}(x)$, each of the eigenvalues $\mathcal{E}_n(k)$, $n \in \Z$, is at least
twofold degenerate, as shown in \cite{Ma}. Thus, even when considering the projector
$P(k)$ corresponding to a single Bloch-Dirac band, one has to deal with the case $m
=2$. This example illustrates the need of the general results stated in
Theorem~\ref{Th main} and Theorem~\ref{Th analytic}.

As for time-reversal symmetry, one checks directly that
\begin{equation} \label{Dirac time reversal}
H_{\rm D}(k)\, T = T \, H_{\rm D}(-k)
\end{equation}
where we introduced the antiunitary operator
\[
T = -i \(1 \otimes \alpha_1 \alpha_3 \) C,
\]
with $C$ denoting complex conjugation in $\Hf$. It is easy to check that $T^2 = 1$
by using the fact that  $\alpha_1 \alpha_3 = - \alpha_3 \alpha_1$.

Let $P(k)$  be the spectral projector  of $H_{\rm D}(k)$ corresponding to a set
$\sigma_0(k)$ satisfying (\ref{Gap condition}), and such that $\sigma_0(k + \lambda
) = \sigma_0(k)$  for all  $ \lambda \in \Gamma^*$ and $\sigma_0(-k) = \sigma_0(k) $.  As
in Section 3, one shows that the map $k \mapsto P(k)$ is smooth and
$\tau$-equivariant. Moreover, from (\ref{Dirac time reversal}) and functional
calculus it follows that
\[
P(-k) = T \, P(k) \, T.
\]
Thus the family $\{ P(k) \}_{k \in \R^3}$ satisfies Assumptions $\mathrm{(P_1)}$,
$\mathrm{(P_2)}$ and $\mathrm{(P_3)}$, and therefore Theorem \ref{Th main} ensures
the triviality of the corresponding complex vector bundle, namely the Bloch-Dirac
bundle.


\end{document}